\documentclass{mem}
\usepackage{natbib}\usepackage{txfonts}\usepackage{balance}
\usepackage{graphicx}
\usepackage[breaklinks,dvipdfm]{hyperref}
\idline{83}{1}

\begin{document}

\title{ The X-ray Properties of Cataclysmic Variables
}

   \subtitle{}

\author{
\c{S}. Balman\inst{1} 
          }

  \offprints{\c{S}. Balman}

\institute{
Middle East Technical University, Department of Physics,
Inonu Bulvar{\i}, Ankara, 06531, Turkey
\email{solen@astroa.physics.metu.edu.tr}
}

\authorrunning{\c{S}.  Balman}

\titlerunning{X-ray Properties of CVs}

\abstract{
Cataclysmic Variables (CVs) are a distinct class of X-ray binaries transfering mass from a donar star to a compact star 
accretor, a white dwarf. They constitute a laboratory for accretion physics, mechanisms and disk theory together with
dynamics of outflows and interaction with surrounding medium.
Our understanding of the X-ray emission from CVs has shown substantial improvement in the last decade with the
instruments onboard present X-ray telescopes. We have a better understanding of the boundary layers and the
accretion shocks using high sensitivity instruments and/or grating X-ray spectroscopy
attaining high spectral resolution yielding temperature and velocity diagnostics. In addition,
high time resolution utitilizes accretion mode diagnostics with characterization of the variability
of the gas flow in discs. Here, I briefly review the X-ray properties of CVs with a highlight on
Intermediate Polars, Dwarf Novae and Classical/Recurrent Novae.

}
\maketitle{}

\section{Introduction}

A Cataclysmic Variable (CV) is a close interacting binary
system in which a white dwarf (WD) accretes material from its
late-type low mass main sequence companion.
CVs can be studied in two main classes.
The accretion occurs through an accretion
disc in cases where magnetic field of the WD is weak or nonexistent ( $B$ $<$ 0.01
MG), such systems are referred as nonmagnetic CVs 
characterized by their eruptive behavior (see Warner 1995, Warner 2003 for a review). 
The other class is the
magnetic CVs (MCVs) divided into two sub-classes according to the
degree of synchronization of the binary. Polars
 have 230 MG$>$$B$$>$20 MG where this strong field
causes the accretion flow to directly channel onto the magnetic pole/s
of the WD inhibiting the creation of an accretion disk (see Cropper
1990; Warner 2003) causing the WD rotation to 
synchronize with the binary orbit. The second
class of MCVs are the Intermediate Polars which
have less field strength of 1-20 MG compared with Polars and are thus asynchronous systems
(Patterson 1994; Warner 2003). 
The space density of CVs is (0.5-10)$\times$10$^{-6}$ pc$^{-3}$ calculated 
using the X-ray luminosity function (a truncated power law) 
depending on the limiting X-ray flux (see Pretorius \& Knigge 2011). The 
cumulative luminosity density of CVs in the 2-10 keV band 
is (0.8-1.4)$\times$10$^{27}$ erg s$^{-1}$ M$_{\odot}^{-1}$ (Sazonov et al. 2006, this value is similar in 17-60 keV range).

\section{Magnetic Cataclysmic Variables}

Magnetic CVs constitute about 25\% of the CV population. About
63\% are Polars (P) and 37\% are Intermediate Polars (IP).
IPs have truncated accretion disks and as the accreting material is 
channeled to the
magnetic poles accretion curtains form. Polars channel matter through accretion funnels
to the magnetic poles. The  X-ray emission is 
from  a strong standing shock near  the surface of the WD (Lamb \& Masters 1979, King \& Lasota 1979). 
The post-shock region is heated to 10-95 keV with 
L$_x$ $\le$ a few $\times$10$^{33}$ erg s$^{-1}$  
(Patterson 1994, Hellier 1996, Kuulkers et al. 2006,
de Martino et al. 2008b, Brunschweiger et al. 2009, Yuasa et al. 2010).
A subclass of IPs show soft X-ray component with blackbody emission kT = 30-100 eV  
(Evans \& Hellier 2007, Anzolin et al. 2008, de Martino et al. 2008a, Staude et al. 2008)  
originating from heated WD surface. 
IPs are the hardest X-ray emitters reaching out to 200 keV 
(INTEGRAL/Swift/Suzaku- Barlow et al. 2006, Landi et al. 2009, Brunschweiger et al. 2009). As
long as the dominating  mechanism is a  Bremsstrahlung, T$_{max}$ of the shock yields
a good approximation for the WD mass in MCVs. In general,a multi-temperature optically thin plasma 
of narrow and broad emission lines with low and/or 
high velocities in excess of 1000 km s$^{-1}$ at
electron densities $>$ 10$^{12}$ cm$^{-3}$ are typically detected
(an isobaric cooling flow type plasma emission
or photoionized gas dominated plasma has been observed, e.g., Mukai et al. 2003, Luna et al. 2010). A
compton reflection component at 6.4 keV, Fe K$\alpha$ line, with EW up to 300 eV are, also, 
characteristics of the emission (de Martino et al. 2008b).

IPs are asynchronous with majority P$_{\rm{spin}}$/P$_{\rm{orb}}$ $\le$ 0.1 
(theoretical range 0.01-0.6) (Norton et al. 2004, 2008;  Scaringi et al. 2010).
Polars show  strong circular and
linear polarization modulated at the binary period (see Beuermann 2004 for an overview).
Polars, also, show asynchronicity at $\le$2\% .
They have mostly short orbital periods below the period gap.
Polars have a dominant soft X-ray blackbody component with kT$\simeq$10-30 eV (Mauche et al. 1999, Traulsen et al. 2011)
due to blobby accretion and WD heating where only about 33\% has no soft component.
Cyclotron cooling  dominates the Bremstrahlung cooling as  the magnetic field increases
and the hard X-ray tails appear at low field strengths for Polars along with soft X-ray excess
increasing with magnetic field strength.
Low accretion rate Polars have L$\le$ $\times$10$^{30}$ erg s$^{-1}$. 
Low and high states occur due to nonuniform accretion and in any survey 50\% 
of Polars are in a low state (Ramsay et al. 2004).
MCVs show energy dependent X-ray/UV/optical pulses  and 
harder/softer poles for a given source. MCVs show the orbital ($\Omega$) and spin ($w$) periods
in their X-ray power sprectra with appereance of sidebands 
(e.g., 2$w$-$\Omega$).
The power spectra of the X-ray light curves are good accretion mode diagnostics  
with a peak at $w$ yielding disc fed,   
peaks at $w$, $w$-$\Omega$ yieding  stream-fed accretion, together with 
$w$, $w$-$\Omega$ and $\Omega$ appearing when  disc overflow occurs as well.
In addition, IPs show orbital modulations (Parker et al. 2005) which has been
suggested as an effect of absorption on the binary plane.

\section{Non-magnetic Cataclysmic Variables}

In Non-magnetic CVs, the  WD accretes matter from late-type MS star filling the Roche Lobe
forming an accretion disk around the WD reaching all the way to the WD since the
magntic field is very low or non-existent (see Warner 1995, 2003). There are several subclasses: (1)
Dwarf Novae inwhich matter is transfered at continuous or sporadic rates, accretion is 
interrupted every few weeks to months by intense accretion (outburst) of days to weeks with
10$^{39}$-10$^{40}$ erg involved energy and the optical magnitudes change by $\Delta m$=2-6; (2)
Nova-like  Variables with high accretion rates (3) the Classical and Recurrent Nova where
explosive ignition (Thermonuclear Runaway) of accreted matrial on the surface of
the WD occurs and material is ejected from the CV system with 10$^{43}$-10$^{46}$ erg of energy 
($\Delta m$=9-10 magnitudes in the optical wavelengths).

The material in the inner disc of Non-magnetic CVs initially moving with Keplerian velocity dissipates its
kinetic energy in order to accrete onto the slowly rotating WD creating a boundary layer (BL)
, the transition region between the disc and the WD (see Mauche 1997 for an overview).
Standard accretion disc theory predicts half of the accretion luminosity to originate
from the disc in the optical and ultraviolet (UV) wavelengths and the other half to emerge from the
boundary layer as X-ray and extreme UV (EUV) emission which may be summerized as
L$_{BL}$$\sim$L$_{disk}$=GM$_{WD}$$\dot M_{acc}$/2R$_{WD}$=L$_{acc}$ (Lynden-Bell \& Pringle 1974, 
Godon et al. 1995). During low-mass accretion rates, $\dot M_{acc}$$<$10$^{-(9-9.5)}$M$_{\odot}$, 
the boundary layer is optically 
thin (Narayan \& Popham 1993, Popham 1999) emitting mostly in the hard X-rays (kT$\sim$10$^{(7.5-8.5)}$ K).
For high accretion rates , $\dot M_{acc}$$\ge$10$^{-(9-9.5)}$M$_{\odot}$, the boundary layer is expected to be
opticallly thick (Popham \& Narayan 1995) emitting in the soft X-rays or EUV (kT$\sim$10$^{(5-5.6)}$ K). 
Since the non-magnetic CVs are diverse, here I concentrate on  dwarf nova and classical nova systems for further 
X-ray observations.
 
\subsection{Dwarf Novae}

During quiescence (low-mass accretion rates) of DN systems the boundary layer is  detected in the hard X-rays.
The X-ray spectra are mainly characterised with a multi-temperature isobaric cooling flow type
model of plasma emission at T$_{max}$=9-55 keV with accretion rates of 10$^{-12}$-10$^{-10}$ M$_{\odot}$ yr$^{-1}$.
The X-ray line spectroscopy indicates narrow emission lines (brightest OVIII K$\alpha$) 
and near solar abundances, with a 6.4 keV line due to reflection from the surface of the WD. 
The detected Doppler broadening in lines during quiescence is $<$750 km s$^{-1}$ with 
electron densities $>$10$^{12}$ cm$^{-3}$ 
(see  Perna 2003, Baskill et al. 2005, Kuulkers et al. 2006, Rana et al. 2006, Pandel et al. 2005, 
Balman et al. 2011).
A missing BL in the X-rays have been suggested due to low L$_x$/L$_{disk}$ ratio. However, it has been
pointed out that BL may emit significant fraction of its luminosity in the EUV/FUV close to the star 
(Sion et al. 1966, Pandel et al. 2005).

DN outbursts are brightenings of the accresion discs as a result of thermal-viscous instabilities summerized
in the DIM model (Disc Instability Model; Lasota 2001,2004).
During the outburst stage, DN X-ray spectra differ from the quiescence since the
accretion rates are higher (about two orders of magnitude), the BL is expected to be optically thick emitting
EUV/soft X-rays (Lasota 2001, see the X-ray review in Kuulkers et al. 2006).
Soft X-ray/EUV temperatures are in a range 5-30 eV are detected from some systems (e.g., Mauche et al. 1995, Mauche \&
Raymond 2000). As a second emission
component, DN show hard X-ray emission during the outburst stage however, at a lower flux level and X-ray
temperature compared with the quiescence (e.g., WW Cet \& SU UMa: Fertig et al. 2011, SS Cyg: MacGowen et al. 2004).
On the other hand, some DN show increased level of X-ray emission including soft X-rays 
(GW Lib \& U Gem: Byckling et al. 2009). The grating spectroscopy of the outburst data indicate 
large widths for lines with velocities in excess of 1000 km s$^{-1}$ (Mauche 2004, Rana et al. 2006).

In the accretion process the material travels from outside inwards and any time variations of
the mass transfer rate in the flow is also transported inwards. This self-similar variability
of accretion rate is refered as flicker noise (Lyubarskii 1997; see also Anzolin et al. 2010
for a review for CVs). Low frequency perturbations  generated at the outer disc
propogate towards the WD and finally to the X-ray emitting region. Variations occur at any
radii on dynamic  timescales and most variability emerges from inner regions (Churazov et al. 2001).
Thus, studying the broad-band noise characteristics of the power spectra from cataclysmic variables
can shed light on the inner disk geometry and dynamics. Disk truncation radii
for IP systems have been observationally detected this way (R$_{in}$=1.9$\times$10$^{9}$ cm, 
Revnivtsev et al. 2011)
The missing BL as proposed from earlier results and UV/X-ray delays detected in the DN outbursts
are indication of possible disk truncation in DN. Truncated accretion disks in Dwarf Novae may also
help to explain quiescent brightness levels and also high
$\dot M$ during quiescence in some DNs.
Models have  been suggested where  irradiation by the WD (King 1997), a coronal
siphon flow (Meyer  \& Meyer-Hofmeister 1994, Liu et al. 1995), WDs rotating near break up velocity (Ponman
 et al. 1995), a spherical corona (Mahasena \& Osaki 1999), or a 
hot settling flow  (Medvedev \& Menou 2002) that may be responsible for the 
distuption of the  inner disc explaining the lack of X-ray emission and formation of coronal flows.

\subsection{Classical and Recurrent Novae}

Classical novae (CNe) outbursts occur in Cataclysmic Variable (CV) systems
on the surface of the WD as a result of an explosive ignition of
accreted material (Thermonuclear Runaway-TNR) causing
the ejection of 10$^{-7}$ to 10$^{-3}$ M$_{\odot}$ of material at velocities
up to several thousand kilometers per second (Shara 1989; Livio 1994; Starrfield 2001; Bode \& Evans 2008).
The classical nova systems have an initial low level accretion phase 
($\le$10$^{-11}$ M$_{\odot}$) where the recurrent nova generally show 
higher accretion rates. Once the critical pressure at the base of the
WD envolope is reached (e.g., 10$^{19}$ dyn cm$^{-2}$) the temperatures reach T$\sim$10$^8$ K, a TNR occurs and the 
hydrogen on the surface of the WD starts burning. Next, the inverse Beta-decay 
reactions produce isotopes that reach the outer layers of the convective envelope 
dumping enough energy expanding the outer layers. In the initial
expansion phase the WD envelope reaches to sizes afew $\times$10$^{12}$ cm
and a visual maximum is attained.
A gradual hardening of the
stellar remnant spectrum with time past visual maximum is expected consistent with H-burning at 
constant bolometric luminosity and
decreasing photospheric radius, as the envelope mass is depleted. 
This residual
hydrogen-rich envelope matter is consumed by H-burning and wind-driven
mass loss. Some of the important parameters that govern nova evolution are
the initial accretion rate, the age of the white dwarf,
the mass of the white dwarf and the 
composition of the envelope over the white dwarf surface ( 
Prialnik \& Kovetz 1995; Starrfield et al. 1998; Yaron et al. 2005).

The emission from the remnant WD is mainly a blackbody-like stellar continuum
refered as the soft X-ray component.
As the stellar photospheric radius decreases in time during the constant bolometric
luminosity phase, the
effective photospheric temperature increases
(up to values in the range 1--10 $\times 10^5$ K) and the peak of
the stellar spectrum is shifted from visual to ultraviolet and
to the X-ray energy band (0.1-1.0 keV), where finally the H-burning turns off
(Balman et al. 1998; Balman \&\ Krautter 2001;
Orio et al. 2002; Ness et al. 2007; Nelson et al. 2008; Page et al. 2010;
Rauch et al. 2010, Ness et al. 2011).

CNe is, also, detected in the hard X-rays (above 0.5 keV) as a
result of shocked plasma emission
{\it during the outburst stage} referred as the hard X-ray component.
The main mechanisms responsible for this component are : (1)
circumstellar interaction  
(Balman 2005,2010,  Bode et al. 2006, Sokoloski et al. 2006,); 
(2)  wind-wind interaction
(Mukai \& Ishida 2001, Orio et al. 2005; 
Lynch et al. 2008, Ness et al. 2009); (3) stellar wind instabilities and X-ray 
emission (as in Owocki \& Cohen 1999); (4) mass accretion
and flickering X-ray light curves, 6.4  keV  Fe lines  
(Hernanz \& Sala 2002, 2007; Page et al. 2010).
Comptonized X-ray emission from the Gamma-rays produced in radioactive decays      
after the TNR has been suggested as a possible hard X-ray Component 
(example:  22Na, 7Be, 26Al; Hernanz et al. 2002).

For the soft X-ray component, originally 
spectral analysis were done using interstellar absorption (hydrogen column density)
and blackbody emission models which yielded super-Eddington luminosities
for the stellar remnant. To correct this, atmosphere models of
C-O and O-Ne core WDs with LTE (Local Thermodynamic Equilibrium)
were used along with interstellar absorption with the $ROSAT$ and $Beppo$-$Sax$ data  
(e.g, Balman, Krautter, \"Ogelman 1998,
Balman \& Krautter 2001).
After the X-ray grating data were obtained, detailed emission and absorption features
were detected with present observatories. Soft X-ray spectra were fitted
using hydrostatic 
NLTE atmosphere models along with interstellar absorption by equivalent hydrogen column density
(Orio et al. 2002, Nelson et al. 2008, Rauch et al. 2010, Osborne et al. 2011, Ness et al. 2011) that would account for
the absorption edges/lines from a static atmosphere. However, the detailed structure 
in the spectra was very difficult to model
yielding best approximations to the observed spectra with large
$\chi^2$ values.
In addition, most absorption features showed blue-shifts in the spectra
(e.g. Ness et al. 2007, 2009) not modeled by the NLTE static atmosphere models.
In order to compensate for this, stellar atmosphere code PHOENIX
(Hauschildt \& Baron 2004)
was adjusted to model NLTE expanding atmosphere models, a hybrid atmosphere model
that is hydrostatic at the base with an expanding envelope on top. These models have been 
used to fit data with yet again best approximations
yielding estimated spectral parameters
(e.g. Petz et al. 2005, van Rossum \& Ness 2010). 

\section{Some Recent Results and Future Prospects}

In the previous\ sections, I reviewed a general perspective
on observational X-ray  properties of Cataclysmic Variables.
In this section, I discuss some current research prospects relating
to X-ray emission from CVs and future work.

One of the important issues of the MCVs is the modeling
of the accretion shock and the column structure 
(i.e., density/temperature; also, stratification) 
and its possible variations 
(see Yuasa et al. 2010). 
Variations on the maximum temperature attained and the derived WD masses indicate
that the structure in the columns may be changing. Recent work on EX Hya by Pek\"on
 \& Balman 2011 reveals that the cold absorption from the bulge at the accretion impact zone
and the neutral column density in the accretion column along with the
covering fraction seems to be indirectly proportional. 
Figure 1 top panel shows the orbital phase-resolved X-ray spectra of EX Hya.
In general, results indicate that as the size of the bulge gets larger, the accretion
curtain size and the absorption is getting smaller. 
Moreover, the maximum temperature gets hotter as the
absorption and the covering fraction in the accretion curtain gets smaller. 
In general, such changes in the
intermediate polar sources suggests that shock structure in the column  
is not persistent and the accretion geometry in the 
system is also variable. This may relate to small changes on the accretion 
rate and the small variations in the
magnetic field magnitude and geometry. It is possible that this is a general 
feature of the MCVs and such variations on the accretion geometry, the column 
temperature changes and field variations can be followed well by monitoring observations 
using simultaneous orbital and spin phase-resolved spectroscopy of these systems. 

\begin{figure}
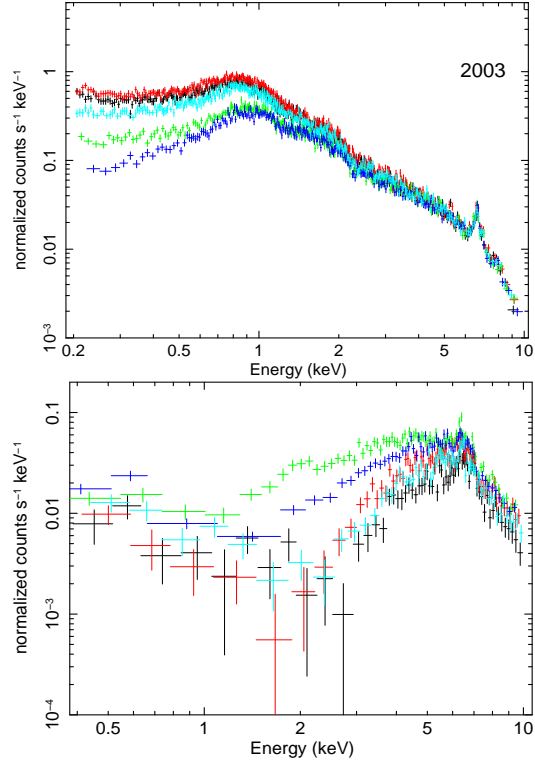

\includegraphics[width=5cm,height=7cm,angle=-90]{ex_2003.ps}
\includegraphics[width=5cm,height=7cm,angle=-90]{fo_orbit_sp2.ps}
\caption{The top figure shows the orbital-phase resolved spectra of EX Hya in 2003.
The bottom figure is the orbital-phase resolved spectra of FO Aqr in 2004. Details
can be found in Pek\"on \& Balman (2011) and Balman \& Pek\"on (2011a).}
\end{figure}

Another intresting class of objects within the IP class is the
small group that hosts a warm absorber; V1223 Sgr, V709 Cas, and RXJ173021.5-055933
(detected from OVII K-edges).
It is suggested that these absorbers exists in the pre-shock in the 
accretion curtain/column.
However, lately the detection of the warm-absorber in the orbital phase-resolved spectra 
of FO Aqr (see Balman \& Pek\"on 2011a) at the bulge of the accretion impact zone 
reveals that just like Low-mass 
X-ray binary dippers CVs can have warm-absorbers on the disk. Figure 1
bottom panel shows the orbital phase-resolved X-ray spectra of FO Aqr and Figure 2
is the variation of the equivalent neutral hydrogen density over the binary phase.  
The minimum and maximum spectra over the orbital phase have been simultaneously
fitted assuming a simple cold absorber and a $warm\ absorber$  model given the same
plasma emission components yielding Log ($\xi$) = 2.2 ($\xi$=L/nr$^2$)
and N$_{warm}$ = (6.5$\pm$1.6)$\times$10$^{22}$ cm$^{-2}$. 
Present and future X-ray missions with high sensitivity
can be used to reveal the structure on the disk via orbital phase-resolved spectroscopy
and the pre-shock by spin phase-resolved spectroscopy which will reveal if the
X-ray emission from Intermediate Polars and/or Polars are being processed and changed  
in the systems via ionized warm absorption in the X-rays 
and how this in the end effects the long and short
term variation of the shock characteristics.    

\begin{figure}
\includegraphics[width=4.65cm,height=6.8cm,angle=-90]{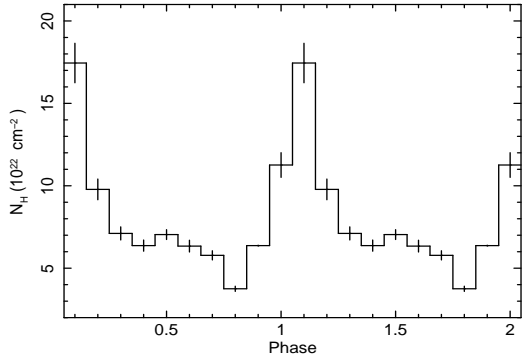}
\caption{
The orbital-phase resolved neutral column density variation
in the spectra FO Aqr in 2004 (Balman \& Pek\"on 2011a).}
\end{figure}

The X-ray emision from boundary layers have
been found to be less than predicted as mentioned before. 
A recent work by 
Balman \& Revnivtsev (2011, in preperation, see also Balman et al. 2011)  show that for five
DN systems, SS Cyg, VW Hyi, RU Peg, WW Cet and T leo, the UV and X-ray power
spectra show breaks in the variability with break frequencies 
indicating the inner disc truncation in these 
systems. The truncation radii for DN are calculated in a range (3-10)$\times$10$^{9}$ cm. 
The authors also analyze the $RXTE$ data of SS Cyg in outburst
and compare it with the power spectral analysis of the
quiescence data where they show that the disc moves towards the white dwarf and receeds
as the outburst declines; between 1-5$\times$10$^{9}$ (see Figure 3) observationally for the
first time. Balman \& Revnivtsev (2011) also calculate the correlation  between
the simultaneous UV and X-ray light curves of five DNe, 
using $XMM$-$Newton$ data obtained in quiescence and find time lags consistent with
delays in the X-rays of 90-200 sec. 
The lags occur such that the UV variations lead X-ray variations
which shows that  as the accreting material travels onto the WD, the variations are
carried from the UV into the X-ray emitting region. 
Therefore, the long time lags of the order of minutes 
can be explained by the travel time of matter from a truncated
inner disk to the white dwarf surface.  The same
study by Balman \& Revnivtsev (2011) (see also Balman et al. 2011)
shows high correlation of X-ray and UV light curves
around zero time lag indicating irradiation effects in these systems. In general, DN
may have large scale truncated accretion disks in quiescence which
can also explain the UV and X-ray delays in the outburst stage and the accretion
may occur through a coronal flow in the disc (e.g., rotating accretion disk coronae).
Note that extended emission and winds are detected from DN in the outburst stage 
which may be an indication of the coronae in these systems (e.g., Mauche 2004). 
It is plausible to monitor DN systems in the X-rays to measure
variability in the light curves in time together with the variations of the disk 
truncation and 
formation of plausible coronae on the disk in quiescence and outburst. 

\begin{figure}
\includegraphics[width=7cm,height=5.5cm,angle=0]{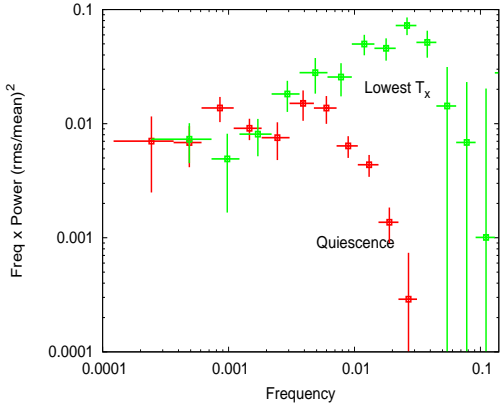}
\caption{
The average Power Spectrum of the SS Cyg in quiescence and outburst when the
X-ray temperatures are the lowest. The entire $RXTE$ data is utilized for the analysis (Balman \& Revnivtsev 2011).}
\end{figure}

The complications in modeling of the soft X-ray component of novae in outburst
have been discussed in the previous section.
Balman \& Pekon (2011b) has used a new approach to analyze
soft component of novae using a complex absorption model 
along with a simple blackbody model for the continuum. 
Main aim has been  to model the absorption components
detected in the high resolution spectra independently from the assumed continuum
model. The complex absorption is of  photospheric , interstellar (hydrogen coumn density)
and of the collisionaly and/or photoionized gas origin in the moving material in the line of
sight from a nova wind or ejecta. The authors utilize photo-ionized warm absorber models and
collisionally ionized hot absorber models for the analysis using the SPEX software
(Kaastra et al. 1996). The results indicate 
blackbody temperatures that are similar
to the expanding NLTE atmosphere model temperatures (for the undelying WD photospheric
temperature).
They model the ionized absorption features simultaneously,  calculating
a global velocity shift for the absorption component in the data originating
from the nova wind/ejecta. Figure 4 shows the fitted $XMM$-$Newton$ RGS data
using collisionally ionized hot absorber model.  
A blue-shifted absorber with  3078-3445 km s$^{-1}$  for V2491 Cyg  and  1085-1603
km s$^{-1}$ for V4743 Sgr have been calculated consistent with wind/ejecta speeds.
They derive CNO abundances from the fits where
V2491 Cyg has a nitrogen overabundance of N=14-36 (ratio to solar abundance)
and C and O are about twice
their solar abundance. V4743 Sgr shows a typical signature of H-burning  with
under-abundant carbon C=0.004-0.2, and enhanced nitrogen N=12-53  and  oxygen O=24-53
(all ratio to solar
abundances). The equivalent hydrogen column density of the ionized absorbers
are (8.0-0.3)$\times$10$^{22}$ cm$^{-2}$ for V2491 Cyg and 
(3.6-7.0)$\times$10$^{23}$ cm$^{-2}$
for V4743 Sgr. The results suggest  that the RGS data of V2491 Cyg is inaccordance 
with both a collisionally ionized absorber (e.g., shocks within winds)
and photo-ionized warm absorber  
modeling major blue-shifted absorption features as the  $\chi_\nu^{2}$ are only sightly
different.
For V4743 Sgr, a photo-ionized warm absorber
(e.g., photo-ionized wind) is a more physically consistent interpretation
with relatively better $\chi_\nu^{2}$ values.

\begin{figure}
\includegraphics[width=5.3cm,height=7cm,angle=-90]{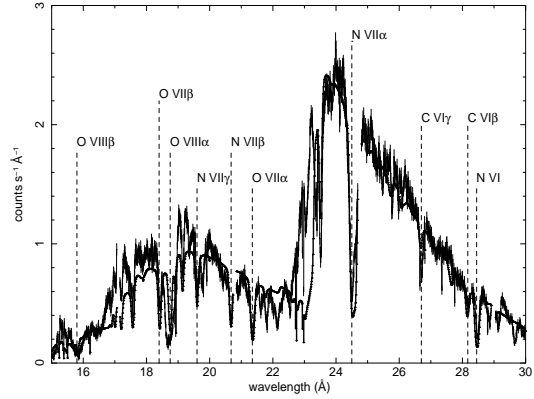}
\caption{
The $XMM$-$Newton$ RGS spectrum of V2491 Cyg fitted with the collisionally ionized hot absorber
model in SPEX. Blue-shifted lines are marked on the figure (see Balman \& Pek\"on 2011b).}
\end{figure}
  
This analysis can be extended to other existing data on these nova and others to see how
complex absorption may have effect on X-ray spectra and how the
absorbers and abundances evolve in time. 
In addition, one can utilize plausible different continuum atmosphere models.  
The systems may show different characteristics depending how deeply embeded the X-ray emitting
region is inside the expending wind/ejecta and where/how much the collisionaly ionized gas 
dominates.

\section{Discussion}


\noindent
{\bf Simone Scaringi: }\ Does the high frequency break move to higher frequencies when looking at higher energies ?\\
\noindent 
{\bf Solen Balman: }\ No not expected. The $RXTE$, $XMM$-$Newton$ EPIC and OM-UV light curves show similar breaks
within errors.\\
\noindent
{\bf Vitaly Neustroev: }\ Where do the orbital modulations in Intermediate Polars come from ?\\
\noindent
{\bf Solen Balman: }\ They are due to cold/warm absorption from the bulge at the accretion stream impact zone
on the outer edge of the disk.

\begin{acknowledgements}
SB acknowledges support
from T\"UB\.ITAK, The Scientific and Technological Research Council
of Turkey,  through project 108T735.
\end{acknowledgements}

\end{document}